\documentclass[11pt]{article}
\usepackage{graphicx}

\newcommand{\ca}{\mathcal{A}}
\newcommand{\cm}{\mathcal{M}}              

\RequirePackage{amssymb}
\usepackage{geometry} 
\begin{document}
     
\begin{center}
{\bf Mass distributions for the Kepler problem} \\[2cm] 
M. Grigorescu \\[3cm]  
\end{center}
\noindent
$\underline{~~~~~~~~~~~~~~~~~~~~~~~~~~~~~~~~~~~~~~~~~~~~~~~~~~~~~~~~
~~~~~~~~~~~~~~~~~~~~~~~~~~~~~~~~~~~~~~~~~~~~~~~~~~~~~~~~~~}$ 
\\[.3cm]
The regularities  in the structure of the planetary system,  expressed  by the Titius-Bode law, can be accounted  by using a more general formula, derived not by fit but from a logarithmic  integrality constraint  on the areolar velocity. This work presents the  elementary adiabatic invariant  used  in constraint, the new formula, and applications  to  the orbit spacing, numbering and mass distributions for  planets and the Jupiter satellites.  \\
$\underline{~~~~~~~~~~~~~~~~~~~~~~~~~~~~~~~~~~~~~~~~~~~~~~~~~~~~~~~~~~~~~
~~~~~~~~~~~~~~~~~~~~~~~~~~~~~~~~~~~~~~~~~~~~~~~~~~~~~}$ 
\vskip9cm
\begin{center}
Dedicated to the anniversary of  \\ 
\underline{450 years since the birth of Johannes Kepler} \\
$\ca nno~~~\cm. ~\cm.~ XXI. $
\end{center}

\newpage

\section{Introduction} 
Regularities observed in the structure of celestial many-body systems, such as the spacings between the planetary orbits, or the spiral branching in the Galaxy, are still a puzzle for the theory  \cite{am}  p. 619, despite the well known expressions of the interactions involved. \\ \indent
In {\it Mysterium Cosmographicum}, (T\"ubingen, 1596), Kepler has tried to relate the intervals between the planetary orbits studied in his time, to the five perfect solids:  tetrahedron (${\cal T}$), cube (${\cal C}$), octahedron (${\cal O}$), dodecahedron (${\cal D}$), icosahedron (${\cal I}$). Thus, each such solid provides a ratio\footnote{Highly regarded by Kepler  was also the "golden ratio" $\tau = 1/(\tau -1) = 1.618...$, related not only to proportions in the polyhedra ${\cal D,I}$ (e.g. L. Pacioli, {\it De Divina Proportione}, (1509)), or artworks, but also to the Fibonacci sequence or the logarithmic spiral.} $\rho = d_M/d_m$, where $d_{M(m)}$  is the maximum (minimum) distance from the surface  to the symmetry center. The ordering $[{\cal O,I,D,T,C}]$ yields a sequence  $[\rho]=[1.73,1.26 ,1.26 ,3,1.73]$, while from the average orbital radii one obtains\footnote{Neglecting the asteroid belt between the orbits of Mars and Jupiter. This belt at $2.9 r_E$ introduces a new interval, and the large ratio 3.42 ($\approx R_4/R_2$ from (\ref{tb})) is in fact the product between  1.91 and 1.79.} $ [\rho_{obs}]=[1.84,1.38 ,1.52,3.42,1.83]$, (or $[1.48,1.33, 1.37,2.97,1.57]$ from Kepler's data on
the "inter-planetary shells") . \\ \indent
A more precise agreement is provided by the empirical rule 
\begin{equation}
R_n =r_E ( 0.4 + 0.3 \cdot  2^n)~~, n=- \infty, 0,1,2,...,8  \label{tb} 
\end{equation}
found by J. K. Titius (1772),  and  re-discovered by J. E. Bode (1776), for the average planetary radius $R_n$, where $n$ is an integer index for the planet, and $R_1=r_E=149.6 \cdot 10^6$ km is the observed value for the Earth. \\ \indent
The Titius-Bode law (\ref{tb}) reproduces well the data for most planets, showing that the orbit spacing is not arbitrary, but it provides no "ordering principle", or indications about suitable  extensions to the periodic motions  in other central fields. The purpose of this work is to discuss 
a more general formula  \cite{qdpr}, derived from an "integrality" constraint on the areolar velocity. The  elementary adiabatic invariant, specific to the Kepler problem, used in this approach, is presented in Section 2. The  integrality  constraint is given in Section 3, with applications extending to the planetary system the previous considerations on  the Jupiter satellites (\cite{qdpr}, Appendix C).  Concluding remarks are summarized  in Section 4.     
\section{ The elementary adiabatic invariant }  
A point-like, nonrelativistic scalar particle of mass $m$, described by the canonical  coordinates  $({\bf q}, {\bf p}) $ on the momentum phase-space  $M= T^*{\mathbb  R}^3$, placed in the central  potential 
$V({\bf q}) = - \eta / \vert {\bf q} \vert$, has the Hamiltonian   
\begin{equation}
 H ({\bf q}, {\bf p}) = \frac{{\bf p}^2}{2m}  -  \frac{\eta }{ \vert {\bf q} \vert}~~. \label{hkep} 
\end{equation} 
With respect to the  Poisson bracket $\{*,* \}$, the dynamical symmetry algebra  ${\mathfrak g}_H$ of $H$, 
$${\mathfrak g}_H = \{  f \in {\cal F}(M) ~;~ \{f, H \}  = 0 \}   $$
is generated by the 3 components of the orbital angular momentum vector  ${\bf L} =   {\bf q} \times {\bf p}$, and 3 components of the Runge-Lenz vector
$ {\bf A} = {\bf L} \times {\bf p} + m \eta {\bf q} / \vert {\bf q} \vert$, such that  \cite{su1} 
$${\bf A}\cdot {\bf L} =0~~,  ~~{\bf A}^2 - 2 m H {\bf L}^2 = m^2 \eta^2 ~~, ~~ \{ {\bf A}, H \} = \{ {\bf L}, H \} =0~~,$$ and
$$ \{ L_i, L_j \} =  \epsilon_{ijk} L_k ~~,~~  \{ L_i, A_j \} =  \epsilon_{ijk} A_k ~~,~~ \{ A_i, A_j \} = - 2 m H  \epsilon_{ijk} L_k ~~.$$
These relations indicate that    
\[  {\mathfrak g}_H   = \left\{ \begin{array}{ll}
\mbox{ ${\mathfrak  so}(4,{\mathbb R}) $} & \mbox{ if  $H <0   $} \\[0.2cm]
\mbox{${\mathfrak  so}(3,1)   $} & \mbox{ if  $H>0 $ ~~.} \\[0.2cm]
\end{array} \right. \]
For an ensemble of  $N$ non-interacting, identical particles, placed in the potential $V$, the 
distribution  function  ${\sf f}$ on $M$, normalized to $N$ \cite{cpw},  satisfies the Liouville equation  
\begin{equation}
\frac{ \partial {\sf f}}{\partial t} + \frac{{\bf p}}{m}  \cdot \nabla
{\sf f} - \nabla V \cdot \nabla_{\bf p} {\sf f} = 0 ~~.                \label{leq0}
\end{equation}
A convenient way to solve this equation  is by using the Fourier transform $\tilde{\sf f}({\bf q},{\bf k},t)$  in momentum,
\begin{equation}
\tilde{\sf f}({\bf q},{\bf k},t) \equiv \int d^3 p ~e^{i{\bf  k} \cdot {\bf p}}{\sf  f}({\bf q},{\bf p},t) ~~. \label{fk1}
\end{equation}
Thus,  if  ${\sf  f}({\bf q},{\bf p},t)$  is a solution of (\ref{leq0}), than its   Fourier transform $\tilde{\sf  f}({\bf q},{\bf k},t)$ satisfies 
\begin{equation}
\partial_t \tilde{\sf f} - \frac{i}{m}  \nabla_k \cdot \nabla  \tilde{\sf f} + i  {\bf k} \cdot (\nabla V) \tilde{\sf f} =0~~. \label{fle}
\end{equation}
Various local observables of interest, such as the particle density ${\sf n}({\bf q},t)$, or the current density ${\bf j}({\bf q},t)$, can be expressed by using  $\tilde{\sf f}$ and its derivatives at   ${\bf k}= 0$, by
\begin{equation}
{\sf n}({\bf q},t) \equiv \int d^3p ~ {\sf f}({\bf q},{\bf p}, t)  = \tilde{\sf f} ({\bf q},0,t)~~,  \label{n}
\end{equation}
\begin{equation}
{\bf j}({\bf q},t)  \equiv  \int d^3p ~ \frac{\bf p}{m} {\sf f}({\bf q},{\bf p}, t)  = - \frac{i}{m} \nabla_k \tilde{\sf f}({\bf q},0, t) ~~. \label{j}
\end{equation}
In general, ${\sf f}({\bf q},{\bf p}, t)$  is specified by the infinite series of partial derivatives of
$\tilde{\sf f}$ with respect to ${\bf k}$ at ${\bf k}=0$. Though, some solutions can be
defined by using only  ${\sf n}({\bf q}, t)$, or a simple functional of ${\sf n}({\bf q}, t)$. In particular, such functionals  remaining unchanged during time evolution will be called coherent. \\
 \indent
An important class of  coherent solutions for (\ref{leq0})  is provided by the   "action distributions"  
\begin{equation}
{\sf f}_0 ({\bf q},{\bf p},t) = {\sf n} ({\bf q} ,t) \delta({\bf  p}- \nabla S({\bf q},t) )~~, \label{cs1}
\end{equation}
remaining all the time a product between  ${\sf n} ({\bf q} ,t)$ and $ \delta({\bf  p}- \nabla S({\bf q},t) )$. The two real functions of coordinates and time, ${\sf n} ({\bf q} ,t)$ and $ S({\bf q},t) $ 
are related by the Hamiltonian flow, because for the Fourier transform   
\begin{equation}
\tilde{\sf f}_0 ({\bf q},{\bf k},t) = {\sf n} ({\bf q} ,t) e^{i {\bf k} \cdot  \nabla S({\bf q},t) )}  ~~,  \label{f0}
\end{equation}
(\ref{fle}) reduces to the coupled equations  \cite{cpw}
\begin{equation}
\partial_t {\sf  n} = - \nabla {\bf  j}   ~~,     \label{co0} 
\end{equation}
\begin{equation}
{\sf n} \nabla [ \partial_t S + \frac{(\nabla S)^2}{2m} +V]=0 ~~, \label{hj}
\end{equation}
where  ${\bf j} \equiv {\sf n} \nabla S /m$ is the current density (\ref{j}). Thus, presuming
the existence of a "momentum potential"  $S({\bf q}, t)$, we get both the continuity  and  the one-particle Hamilton-Jacobi  equations. 
\\ \indent
For the central potential there are  three functions in involution (\cite{am}, p. 301),  $H$, ${\bf L}^2$,  and $L_z$,  such that the Hamiltonian system
\begin{equation}
\dot{\bf q } = \frac{\bf p}{m}~~,~~ \dot{\bf p} = - \frac{\eta}{\vert {\bf q} \vert^3}{\bf q}~~,   \label{kp} 
\end{equation}
is completely integrable\footnote{Some of the angles $\theta_{lm}$ defined by $\cos \theta_{lm} = m/\sqrt{l(l+1)}$  for quantum values of  ${\bf L}^2, L_z$ are also characteristic angles in symmetric polyhedra such as ${\cal C}(l=1)$,  ${\cal O}(l=4)$, ${\cal  D}(l=2,5)$   \cite{gsa}.}.  In the case  $H= E <0$, the integration of  (\ref{hj}) can be reduced to the calculus of orbits for a 4-dimensional harmonic oscillator having the Hamiltonian \cite{geoma}  
\begin{equation}
H_{4d} (\tilde{Q}, \tilde{P})  = \frac{\tilde{P}^2}{2m} + \frac{m \omega^2 \tilde{Q}^2}{2}~~, \label{h4do}   
\end{equation}
where $(\tilde{Q}, \tilde{P})$ are the usual Cartesian coordinates on $T^* {\mathbb R}^4$.
Introducing spherical coordinates $R, \theta, \varphi_1, \varphi_2$ instead of $\tilde{Q}\equiv (Q_0,Q_1,Q_2,Q_3)$, the operator $\nabla_4 \equiv (\partial_0, \nabla_Q )$, $\partial_0 \equiv \partial/ \partial Q_0$,  from the stationary Hamilton-Jacobi equation $H_{4d} (\tilde{Q}, \nabla_4  S ) = \epsilon$, 
\begin{equation}
\frac{1}{2m} ( \nabla_4 S)^2  + \frac{m \omega^2 \tilde{Q}^2}{2} = \epsilon ~~, \label{hjo4} 
\end{equation}
can be written  in the  form
$$ \nabla_4 = {\bf e}_R \frac{ \partial }{\partial R}  + \frac{1}{R} \nabla_{Y4}~~,~~R =\sqrt{ \tilde{Q}^2 } ~~,   $$  
where  ${\bf e}_R \partial_R$,  $ \nabla_{Y4} /R$ are the radial and angular components, respectively. By changing the variable $R$ to  $r=R^2 $, and then dividing (\ref{hjo4}) to $r$, we get
\begin{equation}
\frac{1}{2m} ( 2 \frac{ \partial_r S}{r} {\bf e}_R +   \frac{ \nabla_{Y4} S}{r}  )^2  + \frac{m \omega^2}{2 } = \frac{\epsilon}{r}   ~~. \label{hjo41}
\end{equation}
The angle variables  $\theta, \varphi_1, \varphi_2$  are coordinates on the 3d sphere  ${\rm S}^3 \simeq SU(2) \subset {\mathbb R}^4$, and can be expressed in terms of the  Euler angles $\varphi, \theta, \psi$ for the rigid body \cite{cdr}. Moreover, by taking  $S$ independent of $\psi$, $\nabla_{Y4}S /2 $ reduces to the usual angular part $\nabla_Y S$ of $\nabla S$ in ${\mathbb R}^3$.  In this case, (\ref{hjo41}) becomes 
\begin{equation}
\frac{1}{2m} ( \nabla S )^2   -  \frac{\epsilon}{4r} = -  \frac{m \omega^2}{8 }   ~~,
\end{equation}
which is the same as the Hamilton-Jacobi equation $H({\bf q}, \nabla S)= E$ for the Hamiltonian  (\ref{hkep}), with  
\begin{equation}
\eta =  \frac{\epsilon }{4}  ~~,~~\vert {\bf q} \vert =r~~,~~ E=  -  \frac{m \omega^2}{8 } ~~. \end{equation} 
The ratio $J=\epsilon/ \omega = \eta \sqrt {2m/ -E}$ is an adiabatic invariant, and for a circular orbit in ${\mathbb R}^3$  of radius $r$, one obtains $E_r = - \eta /2r$ and $J_r = 2 \sqrt{m \eta r}= 2 L_r$, where $L=\vert {\bf L} \vert$.  For the gravitational potential the parameter $\eta= m \gamma_G M_o$ depends on the constant $\gamma_G = 6.67 \cdot 10^{-11}$ N$\cdot$m$^2$/kg$^2$ and the mass $M_o$ of the central body (e.g. the Sun for the planets or Jupiter for its satellites).  The particle velocity\footnote{
Given by Kepler's third law, found in the spring of 1618 and published in {\it Harmonices Mundi}, 
(1619).}  $v=\sqrt{\gamma_GM_o /r}$ decreases with $r$, and in principle attains the speed of light $c$ at the (unrealistic small)  radius $r_c=R_G/2$, where $R_G = 2 \gamma_G M_o/c^2$ is the Schwarzschild radius for $M_o$. Thus, $J_c= 2 \sqrt{m \eta r_c}= m c R_G$ can be considered as an elementary adiabatic invariant\footnote{For the Coulomb potential  $V_C= -\eta_C/r$ in the hydrogen atom, $r_c$ is the classical radius of the electron, and $J_c = 2 \alpha \hbar$, where $\alpha=1/137$ is the fine structure constant.}. 
The angular momentum $L_r$ takes the value $J_c$  at $r=2 R_G$.  
\section{ Orbit numbering, spacing, and mass distributions   } 
A many-particle system described by (\ref{leq0}), localized in the volume $V \subset 
{\mathbb R}^3$, has  the angular momentum  
\begin{equation}
{\bf L} =m \int_V d^3q ~ {\bf q} \times {\bf j}({\bf q},t)  ~~,
\end{equation}
where ${\bf j}$ is the current density (\ref{j}). The regularities inspiring  the Kepler's geometric "shell model", or the Titius-Bode law, can be accounted for by presuming that, in certain conditions, a  planar "condensed" state of equilibrium may appear, with massive bodies (planets, satellites), rotating\footnote{Effects of fluctuations in angular momentum due to neutrino emission have been discussed in \cite{hay}.} on orbits selected by the  constraint
\cite{qdpr} 
\begin{equation}
\log_2 (\frac{L_n}{J_c} )^3 = n~~,~~n=0,1,2,...~~,   \label{lqc} 
\end{equation}     
where $L_n=\sqrt{M_p \eta r_n}$ is the orbital angular momentum, and $J_c = M_p c R_G$  the elementary adiabatic invariant, for a body of arbitrary mass $M_p$.  Because $L_n/2M_p$ is the 
areolar velocity,  independent of $M_p$,  (\ref{lqc})  yields for the n'th circular orbit the radius  \cite{qdpr} 
\begin{equation}
r_n = R_G 2^{1+ 2n/3}~,~ n=0,1,2,... ~. \label{rn} 
\end{equation}   
In the  case of Jupiter,  $R_G =2.82$ m, and the (1 bar) "surface" radius $R_J= 71492$ km is between $r_{35 } = 59.6 \cdot 10^3$ km and $r_{36}=  94.5 \cdot 10^3$ km. As indicated in Table 1,  the largest satellites: Io, Europa, Ganimede, Callisto can be associated with the orbital numbers $n=39,40,41,42$,  because only in this case the calculated values ($r_n$) are close to the observed ones, ($r_{obs}$). \\ 

 \noindent
{\bf Table 1.} {\small Comparison between the observed orbital radius ($r_{obs}$) of the Jupiter 
satellites and the calculated value ($r_n$); $M_s$ is the satellite mass.}     \\

\noindent
\begin{tabular}{|c|c|c|c|c|}
\hline
Satellite/n  &   Io/39   &  Eu/40  &  Ga/41 & Ca/42  \\ \hline
 $r_{obs}/10^3$km & 421.6   & 670.8   & 1070  & 1882 \\ \hline
 $r_n/ 10^3$km & 378.5  & 600.8   & 953.7  & 1514 \\ \hline
 $r_{obs} / r_n$ & 1.11   & 1.12   & 1.12  & 1.24 \\ \hline
 $M_s / 10^{20}$kg & 723   & 470   & 1550  & 966 \\ \hline
\end{tabular}
\\[1cm]

For the Sun $R_G= 2.95$ km,  the surface radius $R_S= 6.9 \cdot 10^5$ km 
is between $r_{25} = 6.14 \cdot 10^5$ km and  $r_{26} = 9.7 \cdot 10^5$ km, and the known planets (including also the asteroid ring A), can be assigned to  $35 \le n \le 45$, (Table 2). 
The calculated values $r_n$ are in reasonable agreement with the astronomical data for all planets (e.g. the ratio $r_{obs}/r_n$  is between 0.93 and 1.14), excepting Jupiter, for which $r_n$ is close to $( r_{40}+r_{41})/2$, and it was calculated using $n=40.5$. \\[.5cm] 

 \noindent
{\bf Table 2.} {\small Comparison between the observed average planetary orbital radius ($r_{obs}$), and the calculated values:  ($R_{TB}$)  by the Titius-Bode formula (\ref{tb}), and  ($r_n$)  by (\ref{rn}), using as unit $r_E$; $M_p / M_E$  is the ratio between the mass of the planet and the Earth's mass $M_E$.} \\

\noindent
\begin{tabular}{|c|c|c|c|c|c|c|c|c|c|c|}
\hline
Planet/n  &   $M_e$/35 & V/36 & E/37 & $M_a$/38 & A/39 & J/40,41 & S/42 & U/43 & N/44 &   P/45  \\ \hline
 $r_{obs}/r_E$ & 0.39   & 0.72   & 1  & 1.52  & 2.9 & 5.2 & 9.54 & 19.18 & 30.06 & 39.7 \\ \hline
 $R_{TB}/r_E$ & 0.4   & 0.7   & 1  & 1.6  & 2.8  & 5.2  & 10  & 19.6 & 38.8 & 77.2 \\ \hline
 $r_n/r_E$ & 0.42  & 0.66   & 1.05  & 1.67  & 2.65 & 5.3 & 10.6 & 16.8 & 26.7 & 42.4 \\ \hline
 $r_{obs} / r_n$ & 0.93   & 1.1   & 0.95  & 0.91 & 1.1 & 0.98 & 0.9 & 1.14 & 1.12 & 0.93 \\ \hline
$M_p / M_E$ & 0.055   & 0.85   & 1  & 0.107 & $\sim 0.1$ & 318 & 95.2 & 14.53 & 17.16 & 0.083 \\ \hline
\end{tabular}
\noindent
\\[1cm]

Apart from the scale factor\footnote{Table 1 shows that for Jupiter this factor needs a small  increase, possibly because $R_J > r_{35}$.}  $R_G$, a peculiar common trait for these systems is the mass-weighted  average $\bar{n}$ of the orbital number,
\begin{equation}
\bar {n }= \frac{\sum_n n M_n}{\sum_n M_n} ~~, \label{wa}
\end{equation}
where $M_n$ denotes the mass of the body assigned to the radius $r_n$, ($M_s$ in Table 1 and
$M_p$ in Table 2). Thus, one obtains $\bar{n}=41$ for the planetary system (considering $M_{40}=M_{41}=M_{Jupiter}/2$), and $\bar{n}=40.7$ for the satellites of Table 1. 
\section{Concluding remarks} 
The present attempt to understand the regularities observed in the spacing of planetary orbits is based on the integrality constraint (\ref{lqc}), $L_n = J_c \cdot 2^{n/3}$, where $L_n$ is the angular momentum. By contrast to $\hbar$ in the Bohr quantization condition $L_n = n \hbar$ for the atomic electrons,  $J_c$ depends on mass, and (\ref{lqc}) reduces to the condition  (\ref{rn}) for the orbital radius. Nevertheless, the integer $n$ seems to be relevant for the overall mass distribution\footnote{In atoms the $n$'th atomic shell may contain at most $2n^2$ electrons.} as the weighted average (\ref{wa}) is 41 both for the Jupiter satellites and for the planetary system. Though, the orbits 40,41 around Jupiter carry a satellite each, while the ones around the Sun are "collapsing" into the orbit of Jupiter ($n=40.5$).  Worth noting here is also the large interval of empty orbits ($25<n<35$), between the surface of the Sun and the first visible planet. In the case of Jupiter the orbits $36,37,38$ seem to be empty\footnote{A small satellite, Amalthea, has an orbital radius $\approx 1.2 r_{37}$.}, but one may speculate  that the $16350$ km wide, quasi-stable oval  structure, known as the  Great Red Spot, could evolve\footnote{The longitudinal extent of the oval  decreases, and in 20 years from now it may become circular.}   towards a "preformed satellite", related to the first orbit ($n=35$),  below the 1 bar level. \\

\end{document}